\documentclass[sn-mathphys,Numbered]{sn-jnl}

\usepackage{graphicx}%
\usepackage[running] {lineno}
\usepackage{multirow}%
\usepackage{amsmath,amssymb,amsfonts}%
\usepackage{amsthm}%
\usepackage{mathrsfs}%
\usepackage[title]{appendix}%
\usepackage{color}%
\usepackage{textcomp}%
\usepackage{manyfoot}%
\usepackage{booktabs}%
\usepackage{algorithm}%
\usepackage{algorithmicx}%
\usepackage{algpseudocode}%
\usepackage{enumitem}
\usepackage{listings}%

\begin{document}

\title[Numerical Simulation of the Cleaning Process of Microchannel by an External  Flow]
{Numerical Simulation of the Cleaning Process of Microchannel by an External Flow}


\author*[1,2]{\fnm{Boris S.} \sur{Maryshev}}\email{bmaryshev@mail.com}

\author[2,3]{\fnm{Lyudmila S.} \sur{Klimenko}}\email{lyudmilaklimenko@gmail.com}
\equalcont{These authors contributed equally to this work.}

\affil*[1]{\orgdiv{Department of Control Theory}, \orgname{Lobachevsky State University of Nizhny Novgorod}, \orgaddress{\street{Gagarin Avenue 23}, \city{Nizhny Novgorod}, \postcode{603022}, \state{Nizhegorodskaya Oblast}, \country{Russia}}}

\affil*[2]{\orgdiv{Institute of Continuous Media Mechanics}, \orgname{Ural Branch of RAS}, \orgaddress{\street{Korolev str. 1}, \city{Perm}, \postcode{614990}, \state{Permsky krai}, \country{Russia}}}

\affil[3]{\orgdiv{Institute of Physics and Mathematics}, \orgname{Perm State University}, \orgaddress{\street{Bukireva str. 15}, \city{Perm}, \postcode{614990},  \state{Permsky krai}, \country{Russia}}}


\abstract{The paper is devoted to the study of the process of cleaning a microchannel contaminated by impurity particles settling on its walls. The most common reason of microchannel clogging is the sorption of impurity particles by pore walls or "physical sorption". This paper describes the problem of drift of solid non-interacting particles in a microchannel, which can stick to its walls under the action of the van der Waals forces and break away from the wall due to thermal noise and viscous stresses arising from the flow. The pressure drop is given between the channel inlet and outlet. At the initial moment of time, the channel walls are contaminated with adhered particles, i.e. the walls are uneven, which affects the formation of the flow structure through the channel. Over time, under the action of viscous stresses and thermal noise, the particles break away from the channel walls, causing its cleaning. The interaction of the detached particles with the flow is taken into account in the Stokes approximation. In addition, the model takes into account random particle motion caused by diffusion. The problem is solved numerically within the framework of the random walk model. The evolution of the liquid flow in the channel during its cleaning is obtained: stream function, pressure, and vorticity fields. The dependencies of the volume occupied by settled particles, the flow rate through the channel and the channel gap on time are determined for different values of the interaction force between particles and channel wall. The estimations of the channel cleaning time are obtained and discussed. The possibility of control the cleaning process by the modulation of liquid flux through the channel is investigated. It is shown that resonance phenomena can be observed. The dependencies of cleaning time on the modulation amplitude and frequency is obtained and studied. It is shown that changing the modulation parameters leads to a significant change in the cleaning time and helps control the cleaning process.}


\maketitle

\section{Introduction}
\label{Intro}
Clogging of microchannels due to particle adhesion to the walls is a central problem in many natural and techonogical processes. These are various filtration systems – water purification systems, industrial and household filters used, for example, in catalysis or liquid chromatography. Usually, the service life of the filter element is ultimately limited by the filter clogging time. Another application, which is currently becoming more and more popular, is various microfluidic systems, in which particles are either part of the system or are present in the form of dust or other pollutants. The ability to maintain a flow for a long time dictates the usefulness and service life of microfluidic systems, so particle sorption is very undesirable \cite{flowers2012}. Quite often, in order to extend the service life of both filter and microfluidic systems, they are washed with a sufficiently powerful flow of the working fluid \cite{gothsch2011}.

There are several physical mechanisms that lead to microchannel clogging. The simplest case of clogging is “mechanical” blockage, when particles enter a channel whose characteristic gap is smaller than their own size \cite{frey1999,mays2005}. This approach, which has experimental confirmation, describes the transport of fairly large particles. However, it is known that blockage is observed even when a suspension containing fine particles flows through the channel.

Clogging by fine particles is often explained by the formation of aggregates arising. Such aggregates can be of different structures \cite{Lappa1,Lappa2} a result of interparticle interaction, and then a “new” large particle can lead to mechanical blockage \cite{sharp2005}. Such behavior was observed experimentally in different regions of the channel (near the wall and in the bulk of the liquid) and for particles of different sizes \cite{agbangla2012,gudipaty2011}. Later it was shown \cite{kim2002} that the approach based on the discrete element model (DEM) is more suitable, allowing tracking the motion of each individual particle. Another method that was used to calculate the formation of microparticles aggregates is the numerical method of force coupling \cite{marshall2007}. In this case, the channel clogging mechanism is studied for fairly large particles interacting with each other and with the flow. The authors explain the clogging by mutual aggregation of particles, but the sorption of particles on the wall and the effective narrowing of real channels are not taken into account. However, sometimes it is the narrowing of the channels due to sorption on the wall that turns out to be the determining factor leading to blockage of the channels, which was shown in work \cite{klimenko2020}.

There are also many papers devoted to the study of microfluidic devices \cite{Zhong,Han}. Thus, in \cite{agbangla2014} it was experimentally shown that collective effects, such as the formation of aggregates, do not play an important role in the clogging of the device. One of the first numerical studies of particle deposition was performed in \cite{shahzad2016}. The authors used the adhesion/collision model (JKR), which takes into account the formation and fragmentation of aggregates, as well as the deposition of particles on the channel walls. However, the influence of the channel shape on the flow was not considered, such an influence was taken into account in \cite{klimenko2020}.

The article \cite{klimenko2020} is devoted to the study of the clogging of an initially clean channel by a finely dispersed impurity, taking into account the mutual influence of the shape of the walls (which changes when particles are deposited on them) on the flow of liquid formed in the channel gap under the action of an external pressure difference. In this work, the detachment of particles from the walls is ensured by the action of thermal fluctuations (affecting the force of the particle-wall bond) and viscous forces arising from the flow. The same approach is developed in the present work to study the possibility of cleaning a clogged channel with a flow of clean liquid.
Thus, there are many experimental and numerical studies of microchannel blockage. At the same time, the inverse problem of channel cleaning from finely dispersed impurities has not been practically investigated. Perhaps, its solution is considered to be a trivial generalization of the problem of channel blockage, but, as the present study has shown, cleaning is not always possible and deserved to be studied independently.

The presence of ligth or heavy impurity as a particles leads to convective motion in gravity field \cite{Maryshev1,Maryshev2,Harsha}. Into the microgravity conditions or for close values of particle and carrier liquid denseties this effect becomes tiny  \cite{Cheng} and we do not take it into account in present study.

The paper organized as follows. The formulation of the problem with assumption and governing equations are described in Sec. \ref{ProbState}. In this section the method of solution with numeric algorithm is presented as well. Results of cleaning procedure calculation in terms of integrate characteristics are discussed in Sec. \ref{res}. The last section is conclusion where the main results are summarized.

\section{Problem statement}
\label{ProbState}

To model the cleaning process, we consider a rectangular microchannel, on the walls of which there are "stuck particles" of impurity. At the initial time moment, the channel is clogged by finely dispersed impurities; the particles are held on the walls and among themselves by the van der Waals forces. For channel cleaning from such impurity, the pure liquid is pumped by a pressure drop between the input and output $\Delta P=P_1-P_2>0$. We assume that impurity particles can break away from the channel walls due to the thermal fluctuations (random force) and viscous stress forces acting from the viscous flow. In other words, particles will break away when the sum of stresses (viscous and random) exceeds the attracting force with the wall. The schematic setup of the problem under consideration is shown in Fig. 1. The vertical dimension of the channel is denoted by $H$ and $L$ is the horizontal one.

\begin{figure}
	\includegraphics[width=0.9\linewidth ]{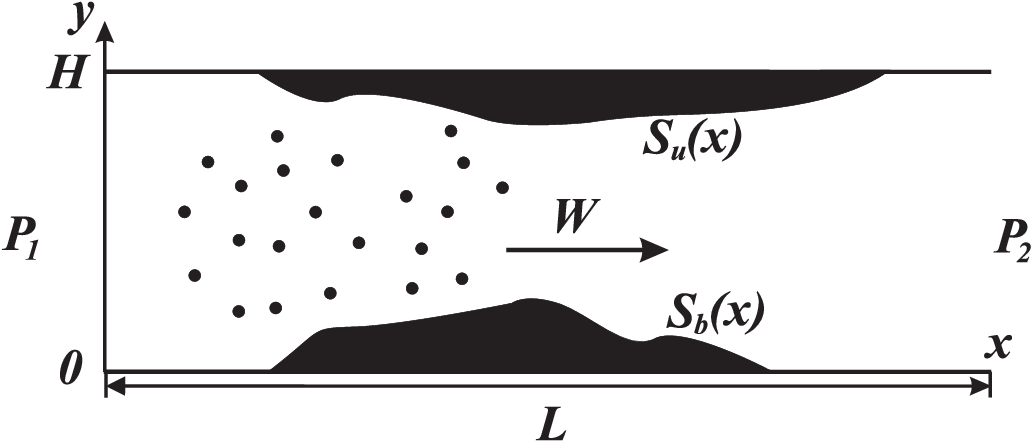} 
	\caption{The sketch of the problem}
	\label{fig1}
\end{figure}

As it is known, inside the clean microchannel (without any impurity) the Hagen-Poiseuille flow is observed

\begin{equation}
	\begin{gathered}
		\mathbf{W}=(u,w)\\
		u=\frac{H\Delta P}{2L\eta}y(H-y),\quad w=0,\\
	\end{gathered}
	\label{Poiselle}
\end{equation}
where $\eta$ is the dynamic viscosity of the liquid. The maximal speed of flow is achieved in the channel center at $y=H/2$ and is denoted by $U=\Delta PH^2/(8L\eta)$. 

To study the flow and impurity transport in a channel with a complex wall shape formed by particles, we apply the model that is developed in \cite{klimenko2020}. The key assumptions of the model are the following.(i)The fluid is incompressible and Newtonian; the fluid flow is laminar and slow; thus, the Reynolds number $Re=HU\rho/\eta\ll 1$ where $\rho$ is the density of the liquid. (ii) The particles are small (with size $a$), they can move together with the flow and be transferred by the random force. The effects of particle inertia are not taken into account due to the low value of the Stokes number $Sk=2Ua^2\rho_p/(9\eta H)\ll 1$ where $\rho_p$ is the impurity density. The random force is generated by the diffusion; thus the Peclet number $Pe=UH/D\sim1$ is moderate. Here $D$ is the diffusivity which is estimated by the Einstein-Smoluchowski relation \cite{einstein1956} as $D= k_B T/(4\pi \eta a)$, where $T$ is temperature and $k_B$ is the Boltzmann constant. (iii) The suspension of impurity particles is dilute, so the mean distance between particles is much greater than the particle size. Thus the interaction between particles will be so rare that particle aggregation can be neglected. Contrary, interactions with the channel walls are significant and are caused by the van der Waals force only. (iv) The densities $\rho$ and $\rho_p$ are comparable or the system is in microgravity conditions for excluding the convective motion.

For the sake of convenience, all governing equations are written in dimensionless form with the following scales for length, time, velocity, and pressure:

\begin{equation}
	H,\quad H/U,\quad (P_1-P_2)H\eta^{-1},\quad \Delta P=P_1-P_2;
	\label{Units}
\end{equation}

The stationary flow of incompressible fluid through the channel in the case of a small Reynolds number can be described by

\begin{equation}
	\begin{gathered}
		\nabla \cdot \mathbf{W}=0\\
		\nabla P=\Delta \mathbf{W}.\\
	\end{gathered}
	\label{base}
\end{equation}

Applying the divergence to the second equation of (\ref{base}), one can obtain the equation for  pressure in the form

\begin{equation}
	\nabla \cdot \left(\nabla P\right)=\Delta P=\Delta \nabla \cdot \mathbf{W}=0,\\
	\label{pres}
\end{equation}
and applying the curl to the same equation gives the equation for the fluid vorticity $\Omega=\nabla \times \mathbf{W}$, specifically

\begin{equation}
	\nabla \times \left(\nabla P\right)=0=\Delta \nabla \times \mathbf{W}=\Delta \Omega.\\
	\label{vort}
\end{equation}

The flow under consideration is two-dimensional one:  $\mathbf{W}=(u,w,0)$  and  $\Omega=\nabla \times \mathbf{W}=(0,0,\phi)$ where 
$\phi=\partial w/\partial x -\partial u/\partial y$. Within these notations, the governing equations (\ref{pres}) and (\ref{vort}) for the flow can be rewritten as 

\begin{equation}
	\begin{gathered}
		\frac{\partial^2 \phi}{\partial x^2}+\frac{\partial^2 \phi}{\partial y^2}=0, \quad 
		\left.\frac{\partial \phi}{\partial x}\right|_{x=0,l}=0,\\
		\frac{\partial^2 P}{\partial x^2}+\frac{\partial^2 P}{\partial y^2}=0, \quad 
		\left.P\right|_{x=0}=1,\quad\left.P\right|_{x=l}=0,\\
		\left.\frac{\partial P}{\partial \mathbf{n}}\right|_{y=S_u,S_b}=0,\\ 
		\left.\frac{\partial \phi}{\partial \mathbf{n}}\right|_{y=S_u}=-\left.\frac{\partial P}{\partial \bf{\tau}}\right|_{y=S_u},\\
		\left.\frac{\partial \phi}{\partial \mathbf{n}}\right|_{y=S_b}=-\left.\frac{\partial P}{\partial \bf{\tau}}\right|_{y=S_b},
	\end{gathered}
	\label{Gover}
\end{equation}
where $S_u(x),S_b(x)$ denote positions of the upper and bottom boundaries of the channel, signs $\partial/\partial \mathbf{n}$ and $\partial/ \partial \bf{\tau}$ are  the normal (the inner normal direction)
and the tangent derivative to the appropriate lines $S_u$ and $S_b$, parameter $l=L/H$ is an aspect ratio of the channel.

The flow velocity can be obtained directly from Eqs. (\ref{base}) as

\begin{equation}
	\begin{gathered}
		\frac{\partial^2 u}{\partial x^2}+\frac{\partial^2 u}{\partial y^2}=\frac{\partial P}{\partial x}, \quad 
		\frac{\partial^2 w}{\partial x^2}+\frac{\partial^2 w}{\partial y^2}=\frac{\partial P}{\partial y},\\
		\left.\frac{\partial u}{\partial y}\right|_{x=0,l}=\left.\phi\right|_{x=0,l}, \quad
		\left.u\right|_{x=S_u,S_b}=0,\\
		\left.w\right|_{x=0,l}=\left.w\right|_{x=S_u,S_b}=0,\\
	\end{gathered}
	\label{Velo}
\end{equation}

Transport of impurity particles is provided by two mechanisms. The first one is a regular movement of particles within the flow; according to an assumption of the low Stokes number ($Sk$) the particle velocity is the same as the fluid flow velocity. The second mechanism is Brownian motion associated with a moderate value of $Pe$. This motion is modelled by the Gaussian random walk. Thus, the particle transport can be described by 

\begin{equation}
	\begin{gathered}
		\delta x_i=u\delta t_i+\sqrt{M(\delta t_i)}f_x,\\
		\delta y_i=w\delta t_i+\sqrt{M(\delta t_i)}f_y, 
	\end{gathered}
	\label{P_move}
\end{equation}
where $i$ is the time step number, $\delta t_i$ is the time step interval, $\delta x_i$ and $\delta y_i$ are the displacements of the particle in the $x$ and $y$ directions per time step, $M$ is the variance of the particle random walk step, and $f_x$, $f_y$ are independent normally distributed random variables (with unit variance and zero mean). The flow velocity field $(u,w)$ can be obtained from the solution of the problem (\ref{Velo}). During the calculations, we use a fixed equal time interval for any steps and choose it on the basis of a variance $M$. According to \cite{einstein1956}, the variance for Brownian motion is $M=2k_BT\delta t/(3\eta\pi a)$. In order to improve the accuracy of calculations, we assume that $M=a^2$ with the smallest space scale of the considered problem. Thus, as a result, the dimension time interval is $\delta t=3\eta\pi a^3/(2k_BT)$. Using the scales (\ref{Units}), one can obtain that $\delta t=3\pi(P_1-P_2) a^3/(2k_BT)$ (see \cite{martens2013}). The current particle position is defined by the summation of all displacements as

\begin{equation}
	\begin{gathered}
		x_n=x_0+\sum_{i=0}^{i=n}\left[u\left(x_i,y_i,t_i\right)\delta t+hf_x(t_i)\right],\\
		y_n=y_0+\sum_{i=0}^{i=n}\left[w\left(x_i,y_i,t_i\right)\delta t+hf_y(t_i)\right],\\
		t_n=n\delta t,
	\end{gathered}
	\label{P_move1}
\end{equation}
where $t_n$ is the dimensionless time moment after $n$ time steps, $h=a/H$ is a dimensionless space step of particle random walk,  $x_0$ and $y_0$ are the initial particle positions. It is natural to choose $x_0=0$ or the inlet of the channel as the initial particle position in the horizontal direction, while the $y$ position is defined randomly inside the interval $y_0=0.2..0.8$. This choice of $y_0$ helps to avoid an immediate interaction with the channel walls, so the injection of particles is performed into the central part of the channel. During calculations of Eqs.\ref{P_move1} if the particle approaches the channel walls, it can stick to the wall due to the van der Waals interaction. The energy of such interaction \cite{elimelech2013} is defined by the expression

\begin{equation}
	E_v=\frac{A}{6}\left[\frac{a}{d}+\frac{a}{d+2a}+\ln\left(\frac{d}{d+2a}\right)\right],
	\label{Energ}
\end{equation}
where $d \gtrsim a$ is the distance from a particle to the channel wall and $A$ is the Hamaker constant, which value belongs to the interval $A \sim 10^{-19} \div 10^{-21} J$ for different substances of channel wall and particle. The value van der Waals force can be estimated by the gradient of energy with respect to distance as

\begin{equation}
	F_v=\frac{\partial E_v}{\partial d}= \frac{A}{3}\frac{2a^3}{d^2\left(d+2a\right)}.
	\label{ForceV}
\end{equation}

As it is expected, Brownian motion of particles prevent the sticking to the wall. According to \cite{ghosh2012}, such random force can be estimated as
\begin{equation}
	F_r=6\pi\eta a V_T=\frac{3}{2}k_BT\sqrt{\frac{2}{aH}},
	\label{ForceR}
\end{equation}
where $V_T$ is the particle thermal velocity \cite{ghosh2012}, $V_T=D\sqrt{2/aH}$, where $D= k_B T/(4\pi \eta a)$ is the diffusivity of Brownian motion.
Comparing these two forces gives the critical distance of particle attachment to the wall;

\begin{equation}
	F_r=F_v(d)\quad \Rightarrow d_c=a\left(\sqrt{1+\frac{2\sqrt{2}AH}{9k_BTa}}-1\right).
	\label{dist}
\end{equation}
If the particle approaches this distance $d_c$, it will be attracted to the wall. As follows from Eq. (\ref{dist}), the critical distance depends on the Hamaker constant, the temperature, and the ratio between particle size and channel size. In the present work we use the distance $d_c\in [a,4a]$, the estimation of its value will be discussed below. 

The present model takes into account two mechanisms of particle detachment: (i) the viscous force generated by the fluid flow and (ii) the random force induced by the thermal noise. The influence of viscosity on the detachment process can be evaluated in terms of the tangent component of the viscous stress tensor \cite{landau1987}:  $\sigma^*_u=\eta\partial u^*/\partial \mathbf{n}|_{x=S_u}$ for the top wall and $\sigma^*_b=\eta\partial u^*/\partial \mathbf{n}|_{x=S_b}$ for the bottom wall, where $u^*$ is the $x$ component of the dimensional flow velocity. In dimensionless units it is easily rewritten as $\sigma=\partial u/\partial \mathbf{n}$.  

To find the dimensionless form of the stresses caused by the random force, it should be noted that unit of stress tensor is identical to pressure unit, specifically for (\ref{Units}), it corresponds to $P_1-P_2$. Due to that fact, to find the dimensionless form of stress tensor  $\sigma$ one should compare it with the force divided to the square of particle size and to the $P_1-P_2$ as the pressure scale . Thus, for the random force in the form of (\ref{ForceR}) one can obtain the dimensionless stress which is produced by the random force as 

\begin{equation}
	\sigma_t=\frac{3k_BT}{\left(P_1-P_2\right)\sqrt{2a^5H}}
	\label{str_T}
\end{equation}

The effect of the particle attachment can be evaluated in terms of the stress as well; for the van der Waals force (\ref{ForceV}) at the distance $d=a$ the corresponding stress component can be written as

\begin{equation}
	\sigma_V=\frac{2A}{\left(P_1-P_2\right)27a^3}.
	\label{str_V}
\end{equation}

Finally, the criterion of particle detachment can be formulated as follows

\begin{equation}
	\begin{gathered}
		\frac{\partial u}{\partial \mathbf{n}}+\sigma_tf>\sigma_V,\\
	\end{gathered}	
	\label{detach}
\end{equation}
where $f$ is the normally distributed random variable (with unit variance and zero mean). 

\begin{figure}[h!]
	\centering{\includegraphics[width=0.3\linewidth ]{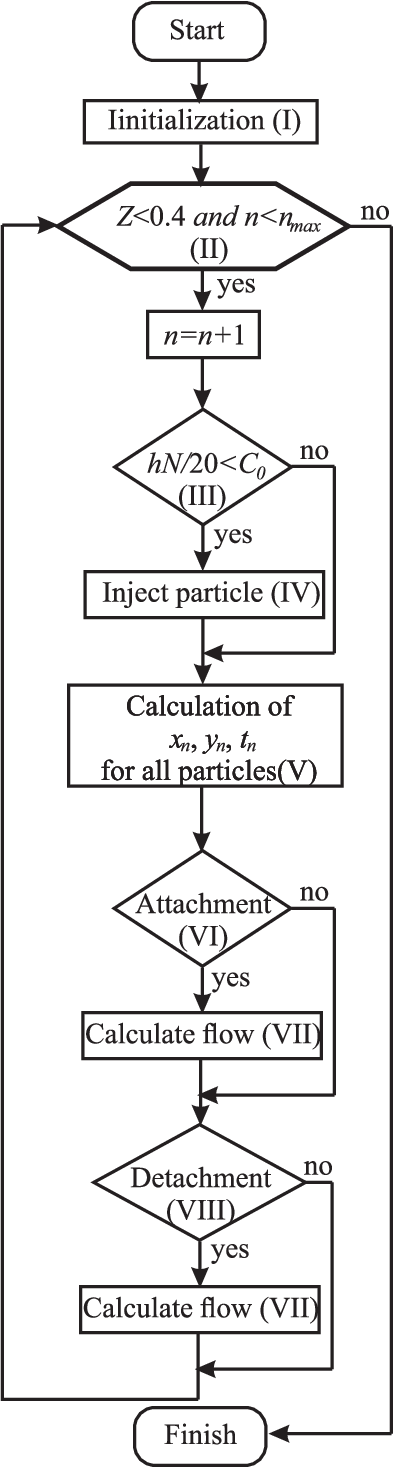} 
	\caption{Algorithm of the channel clogging calculation}}
	\label{Clog}
\end{figure}

The problems (\ref{Gover}) and (\ref{Velo}) were solved for channel clogging and then for channel cleaning. To prepare the microchannel contaminated by impurity particles settling on its walls we have developed a numeric code based on the algorithm  Fig.~\ref{Clog}.
The comments about the procedures of this algorithm is presented below.

\begin{enumerate}[label=\Roman*]
	\item Initially, the channel is filled by the clean viscous fluid, flow velocity corresponds to the Hagen-Poiseuille flow, and $S_b=0$ and $S_u=1$ for all $x\in[0,l]$.
	\item There are two ways of the calculation finishing. The first one, the channel is clogged $Z<0.4$, where $Z=\min_x|S_u(x)-S_b(x)|$ is the minimal channel gap. The second way, the channel can not to be clogged, particle detachment-attachment are in equilibrium, the borders of channel do not change sufficiently:  $n>n_{max}$ and $n_{max}$ is the maximal number of time steps.  Typically, $n_{max}=1000/\delta t$. 
	\item Calculation continues while the initial concentration corresponds to the condition $5h^2N<C_0$. It is natural to assume that the initial volume concentration  reaches value $C_0$  in the region "near" the inlet of channel. The area "near inlet" is estimated as the minimal distance in $x$ direction to the first particle attachment of the wall. The particles are injected at the inlet with $x=0$ and $y\in[0.2,0.8]$ that corresponds to the minimal distance to wall is $\delta y=0.2$. Thus, as particles move with equal probability in $x$ and $y$ directions (see, Eq.(\ref{P_move1})), the minimal distance in the $x$-direction is the same as $\delta y=0.2$. Due to that fact, the total dimensionless volume of zone "near" inlet is $V_n=0.2\times 1$ ($0.2$ in the $x$-direction and $1$ in the $y$-direction for all particle from the inlet). The dimensionless volume, which is occupied by one particle, is $h^2$ and for $N$ particles one can obtain the volume concentration as $C=h^2N/0.2=5h^2N$, where $N$ is the number of particles with $x$ coordinate inside the interval $[0,0.2]$.
	\item The injection of impurity occurs as follows: one by one particle injected into the channel at the inlet ($x=0$), while the $y$ coordinate is a random variable uniformly distributed in the range from $y=0.2$ to $y=0.8$. The injection stops when the volume impurity concentration in the area "near" inlet reaches the  value $C_0$.
	\item The current particles position is calculated from Eqs. (\ref{P_move1}). The particle is excluded from the flow when it attaches to the wall or if the particle reaches the outlet ($x=l$).
	\item Attachment takes place when the $y$-coordinate of the particle reaches the corresponding value. If $y<S_b(x)+d_c$ then the particle attaches to the bottom wall. This particle is excluded from the flow and the bottom wall function becomes $S_b(x)=S_b(x)+h$. If $y<S_u(x)-d_c$  then it attaches to the top wall. This particle is excluded from the flow as well, and the upper wall function becomes $S_u(x)=S_u(x)-h$.
	\item The fluid flow is calculated as a solution of the problems (\ref{Gover}) and (\ref{Velo}) for each event of particle attachment or detachment since the shape of channel walls is changed by these events.
	\item For particle detachment, the condition (\ref{detach}) is being checked for all points at the surface on the both walls (points $\left(x,S_b(x)\right)$ and $\left(x,S_u(x)\right)$ for $x\in[0,l]$, $S_u(x)<1$ or $S_b(x)>0$) . If the particle is detached then it is considered to be re-injected into the flow in the point $(x,S_b(x)+2h)$ for the particle near the bottom wall or in the point $(x,S_u(x)-2h)$ for the top wall.  
\end{enumerate}

\begin{figure}
	\centering{\includegraphics[width=0.3\linewidth ]{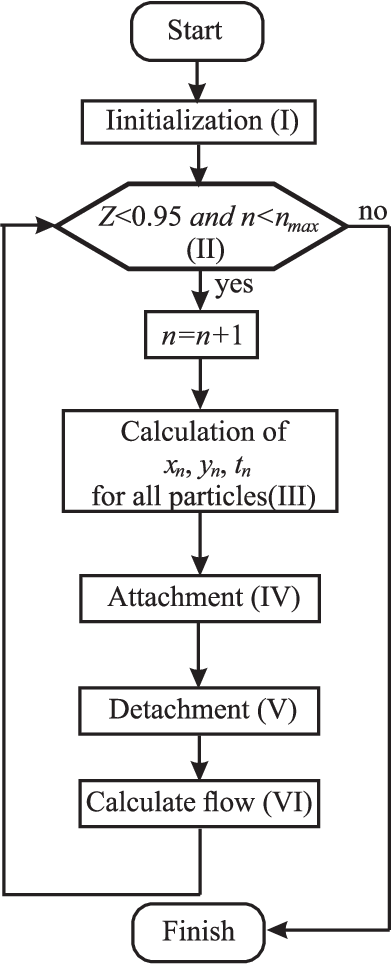} 
	\caption{Algorithm of the cleaning calculation}}
	\label{Clean}
\end{figure}

The algorithm in Fig.~\ref{Clog} is realized, validated and applied for modelling of channel clogging in the paper \cite{klimenko2020}. It is shown that the channel can be clogged only for specific values of problem parameters ($C_0$, $P_1-P_2$, $\sigma_V$, $\sigma_t$). However for any values of ($P_1-P_2$, $\sigma_V$, $\sigma_t$) there can be found some value of concentration $C_0$ when channel is clogged. The aim of present work is to investigate the cleaning process of such clogged channel. For that we prepare the clogged channel by the algorithm Fig.~\ref{Clog} up to the gap $Z=0.4$. After that the cleaning process is calculated by the generally the same algorithm with little simplification (Fig.~\ref{Clean}). It is modelled that the injection of particles to the flow is stopped and channel cleaning by pure liquid and will be discussed below. 

Additionally, we consider the modulation of the pressure drop as the way to control the cleaning process. The modulation leads to the reformulation of the problem (\ref{Gover}) as

\begin{equation}
	\begin{gathered}
		\frac{\partial^2 \phi}{\partial x^2}+\frac{\partial^2 \phi}{\partial y^2}=0, \quad 
		\left.\frac{\partial \phi}{\partial x}\right|_{x=0,l}=0,\\
		\frac{\partial^2 P}{\partial x^2}+\frac{\partial^2 P}{\partial y^2}=0, \quad 
		\left.P\right|_{x=0}=1+A\sin(\omega t), \quad\left.P\right|_{x=l}=0,\\
		\left.\frac{\partial P}{\partial \mathbf{n}}\right|_{y=S_u,S_b}=0,\\ 
		\left.\frac{\partial \phi}{\partial \mathbf{n}}\right|_{y=S_u}=-\left.\frac{\partial P}{\partial \bf{\tau}}\right|_{y=S_u},\\
		\left.\frac{\partial \phi}{\partial \mathbf{n}}\right|_{y=S_b}=-\left.\frac{\partial P}{\partial \bf{\tau}}\right|_{y=S_b},
	\end{gathered}
	\label{Gover_mod}
\end{equation}
where $A=P_a/\Delta P$ is the nondimensional amplitude and $\omega=\Omega L/U$ is the nondimensional frequency of pressure modulation. Here $P_a$ and $\Omega$ are dimensional amplitude and frequency of modulation.

The algorithm for the calculation of cleaning process is presented in Fig.~\ref{Clean}.

\begin{enumerate}[label=\Roman*]
	\item Initially, the clogged channel is prepared by the implementation of algorithm Fig.~\ref{Clog}. 
	\item The calculations finish either when the channel is cleared with $Z>0.95$ ( $Z=\min_x|S_u(x)-S_b(x)|$ is the minimal channel gap) or when $n>n_{max}$ ($n_{max}$ is the maximal number of time steps). In the later case it is assumed that the channel cannot be cleared with the given parameters. Typically, $n_{max}=1000/\delta t$. 
	\item The particles position is determined by calculation of Eqs. (\ref{P_move1}) for all particles in the flow. If particle  reaches the outlet ($x=l$) then it is excluded from the flow. If the particle attaches the wall it is also excluded from the flow.
	\item The particle attachment is almost the same procedure as number (VI) in the clogging algorithm (Fig.~\ref{Clog}), but now there is no need for branching of the algorithm because the flow recalculated at each time step due to changing the boundary conditions. 
	\item The particle detachment is exactly the same procedure as (VIII) in the clogging algorithm (Fig.~\ref{Clog}). 
	\item The flow is calculated by solving the problems (\ref{Gover}) and (\ref{Velo}) for stationary pressure drop or (\ref{Gover_mod}) and (\ref{Velo}) for modulated one. Attachment and detachment of particles usually occurs at each step and the recalculation of the flow is also performed at each step without any "if... then..." constructions. 
\end{enumerate}

The results of cleaning procedure calculation, analysis of cleaning dependencies on the problem parameters are discussed in the next section.

\section{Results and discussions}
\label{res}
Foremost, let us discuss the model parameters that have been used for computation. We consider that the material parameters of the fluid are close to water, i.e., $\rho\approx10^3 kg/m^3$ and $\eta\approx 10^{-3} Pa\cdot s$. The maximal possible size of the particles $a_c$ can be estimated from the assumptions that the Stokes number $Sk\ll 1$ and the Peclet number $Pe\sim 1$, which leads to $Sk\ll Pe$ or $a_c \ll 18\pi H^2 \eta^2/(\rho_pk_b T)$. To find the numerical value of $a_c$, we assume that the densities of fluid and admixture are comparable so $\rho\sim \rho_p$. The maximal value of particle density is $(\rho_p)_{max}=10\rho=10^4 kg/m^3$, the maximal value of temperature is $T_{max}=373 K$, and the minimal possible height of the channel is $H_{min}=1\mu m=10^{-6}m$. Thus, substitution in the ratio above gives $a_c\ll 1m$, which is deliberately done for the considered problem statement. Another restriction for $a_c$ is $a\ll H$; this means that fine particles are considered. We choose the moderate size of the microchannel for microfluidic devices as $H=10^{-5}m$, and the particle size is $a=10^{-7}m$. This size of particle corresponds to very fine particles; however, it is much greater than the molecular size. The length of the channel is greater than $H$, for minimization of computing time, we choose $L=5H=5\cdot 10^{-5} m$. The maximal value of the pressure drop $\Delta P$ can be estimated from the condition for laminar flow $Re\ll1$, thus, as a result, $\delta P_{max}\ll 8L\eta^2/H\rho\approx 400 Pa$. The typical values of the pressure gradient in microfluidic devices are usually in the range $\Delta P/L\in (10^5..10^6) Pa/m$  (see \cite{yang2019,su2010,jin2019}) or, substituting the channel size, for the pressure one can obtain  $\Delta P\in (5..50) Pa$. The range of temperature is $T\in(273,373) K$ and the Hamacker constant for most solid materials is $A\in(10^{-19},10^{-21}) J$ (see \cite{elimelech2013, klimenko2020}). Using Eqs. (\ref{str_T}) and (\ref{str_V}) one can obtain the estimations of stresses in the range $\sigma_T\in(0.016,0.218)$ and $\sigma_V\in(0.0015,1.4815)$. Further, an influence of stresses on the cleaning process is demonstrated.

\begin{figure}
	\includegraphics[width=0.6\linewidth ]{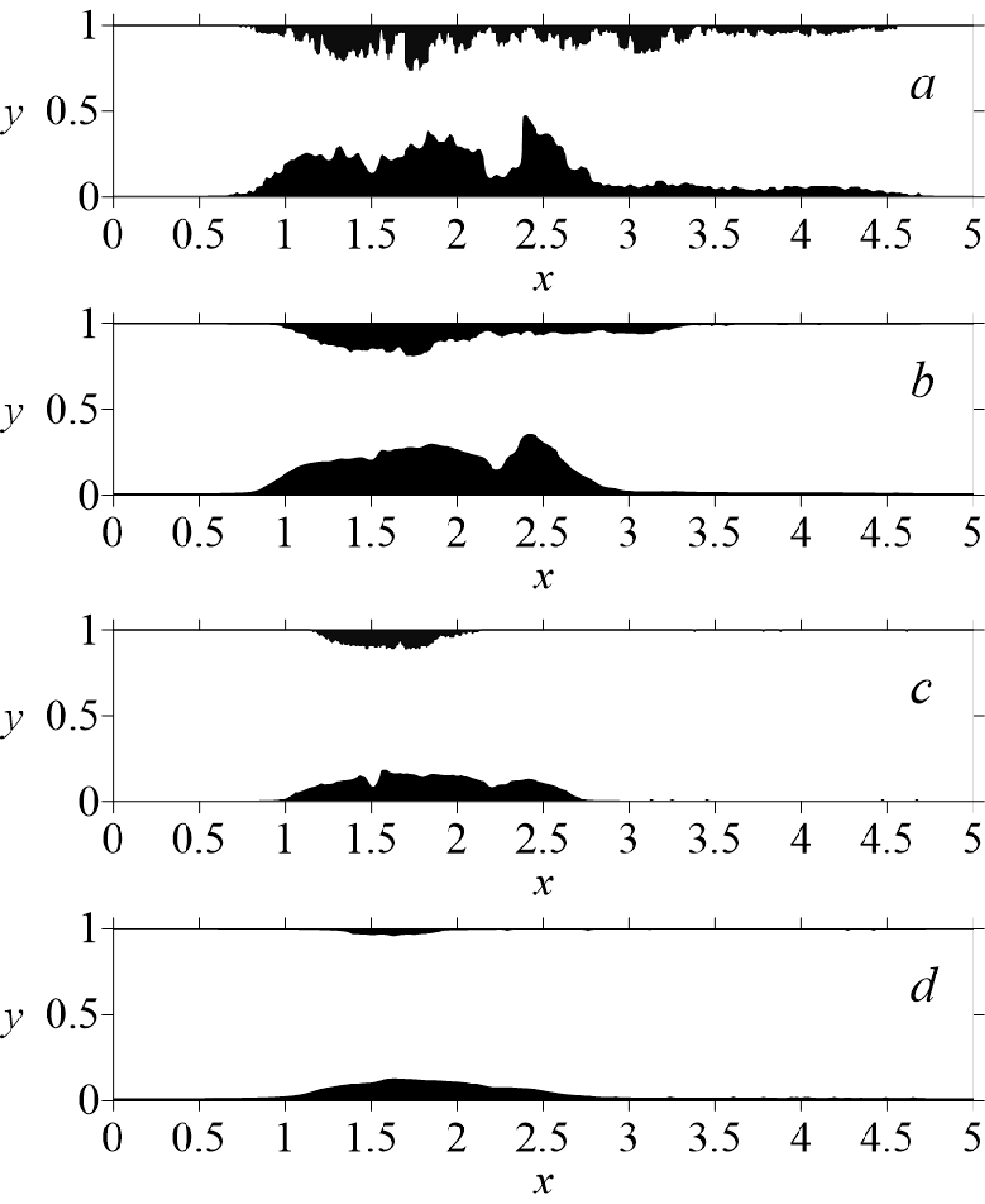} 
	\caption{Position of the walls of the channel contaminated with impurity in different time moments, the area occupied by impurity particles is shaded in black}
	\label{Clean_proc}
\end{figure}

The typical picture of the microchannel cleaning by the steady flow is shown in Fig.~\ref{Clean_proc}. Depending on the problem parameters, the initially clogged microchannel quite fast becomes clean everywhere except for the largest particle deposits (see Fig.~\ref{Clean_proc}). Complete cleaning can be achieved, as will be shown later, in some parameter range; otherwise, the channel remains clogged.

Since the cleaning process is predominantly random, important information can be extracted from the integral parameters. For this purpose we performed 50 independent realizations for each value of the problem parameters. The resulting values of the integral parameters are the arithmetic mean of these realizations. Further, we have analysed the cleaning process based on three integral variables. First one is the volume concentration of the settled impurities, which is determined as $Q=V_p/V$, where $V_p$ is the volume occupied by the settled impurity (black part in Fig. \ref{Clean_proc}), $V$ is the volume of the whole channel space. The second one is the channel gap, that is the minimal distance between the channel walls:

\begin{equation}
	Z=\min_x(S_u(x)-S_b(x))
	\label{gap}
\end{equation}
And the third integral parameter is the fluid flux through the channel ($J$). The flux can be determined by 

\begin{equation}
	J=\int_{y=0}^{y=1}u(x=l,y)dy
	\label{Flux}
\end{equation}

\begin{figure}
	\includegraphics[width=0.7\linewidth ]{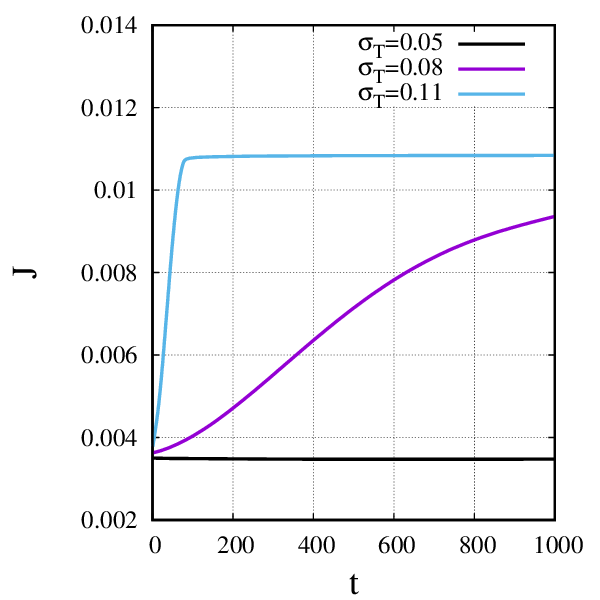} 
	\caption{The dependencies of fluid flux $J$ on time for different values of thermal noise stress $\sigma_t$ which is indicated in legend. The van der Waals force stress value is $\sigma_V=0.3$.}
	\label{J_T_ST}
\end{figure}
\begin{figure}
	\includegraphics[width=0.49\linewidth ]{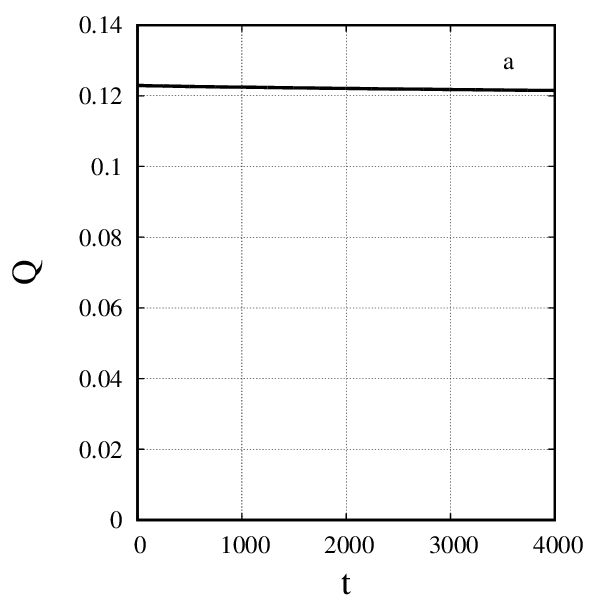} 
	\includegraphics[width=0.49\linewidth ]{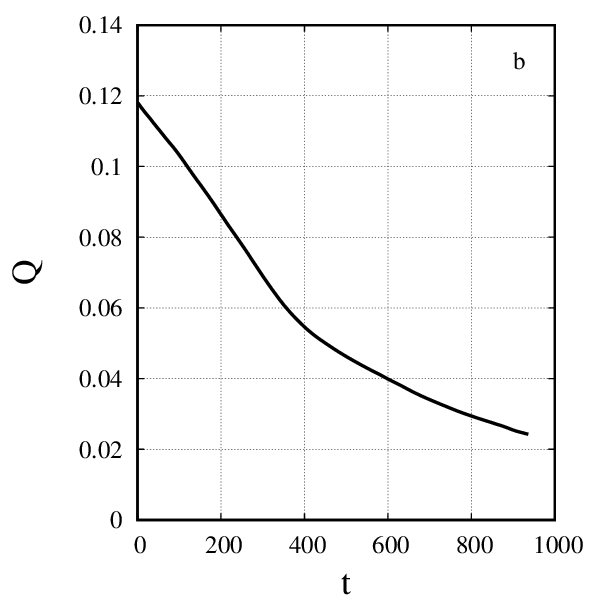} 
	\includegraphics[width=0.49\linewidth ]{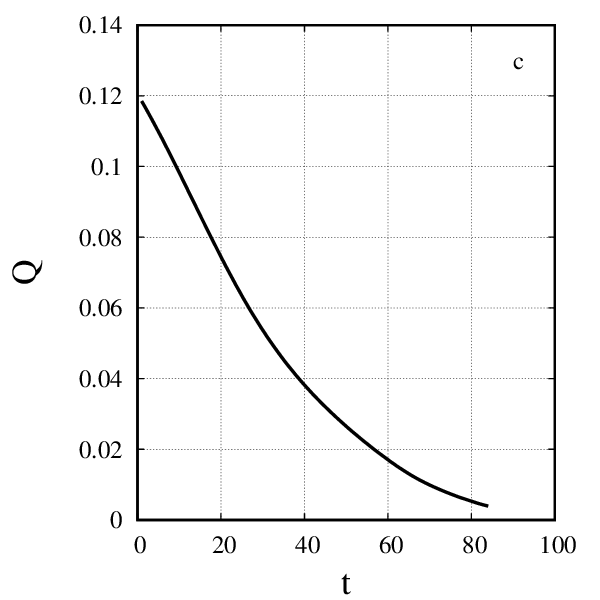} 
	\includegraphics[width=0.49\linewidth ]{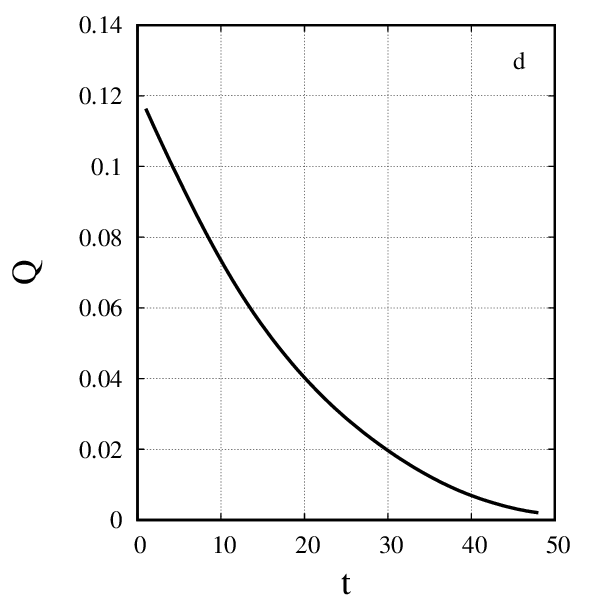} 
	\caption{Dependencies of occupied volume concentration on time for different values of thermal noise stress: 
		a -- $\sigma_t=0.05$,  b -- $\sigma_t=0.08$,  c -- $\sigma_t=0.11$, and  d -- $\sigma_t=0.15$. The van der Waals force stress value is $\sigma_V=0.3$.}
	\label{Q_T_ST}
\end{figure}

Let us start the description of the cleaning process with an analysis of the thermal fluctuations $\sigma_t$ impact for fixed values of the problem parameters, in particular, $\sigma_V=0.3$. Thereby, figure \ref{J_T_ST} shows dependencies of the fluid flux ($J$) through the channel on time. All lines of Fig. \ref{J_T_ST} demonstrate a tendency to flow rate growth as thermal fluctuations $\sigma_t$  increase. However, for small temperature (or small thermal noise), the flux is constant. The channel cannot be cleared, and flux remains such a small value. With increasing the temperature, the cleaning process speeds up. The same trend can be recognized from the dependence of the volume concentration on time. Figure~\ref{Q_T_ST} shows such dependencies for different values of thermal noise stress $\sigma_t$.

From Fig.~\ref{Q_T_ST}a it is clear that for small temperature ($\sigma_t=0.05$) the cleaning process does not occur; the concentration value remains practically unchanged for a long time. With the increasing of $\sigma_t$ the channel cleaning is strongly accelerated. So, for $\sigma_t=0.15$, all particles leave the channel at 50 (dimensionless units of time); thus, the cleaning time is reduced by a factor of 20 compared to $\sigma_t=0.08$. 

\begin{figure}
	\includegraphics[width=0.49\linewidth]{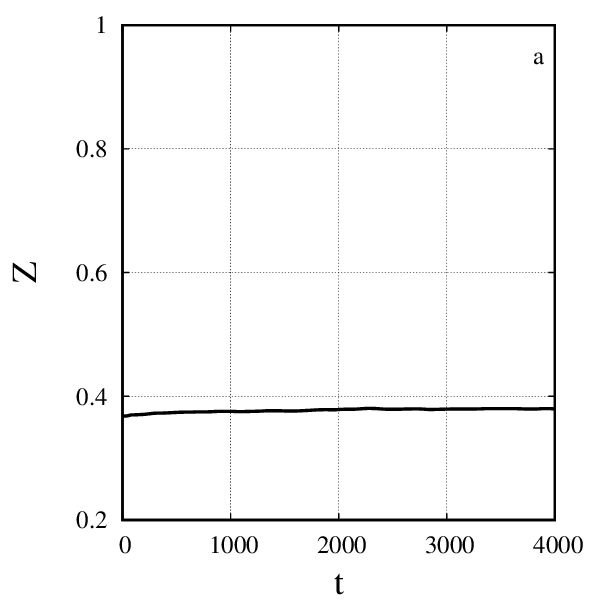} 
	\includegraphics[width=0.49\linewidth]{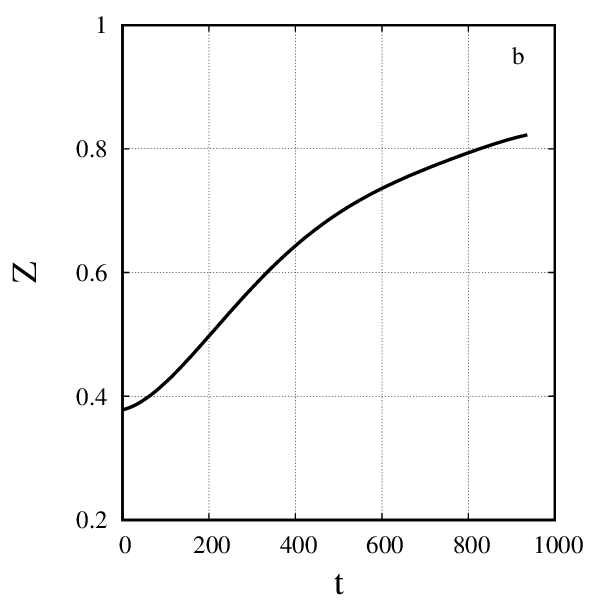} 
	\includegraphics[width=0.49\linewidth]{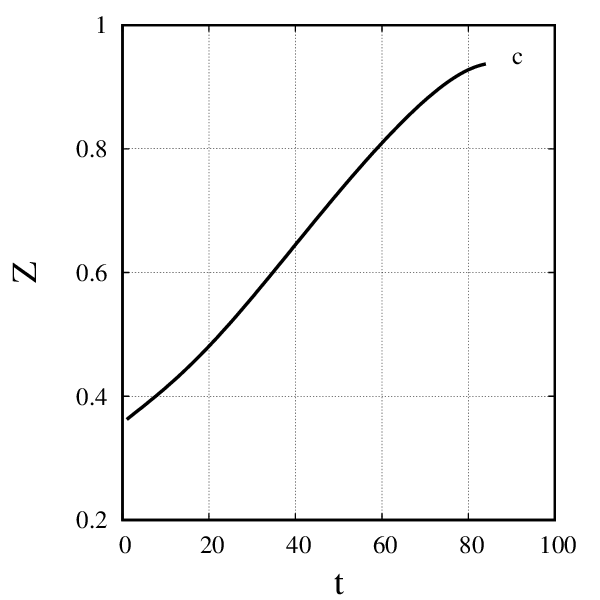} 
	\includegraphics[width=0.49\linewidth]{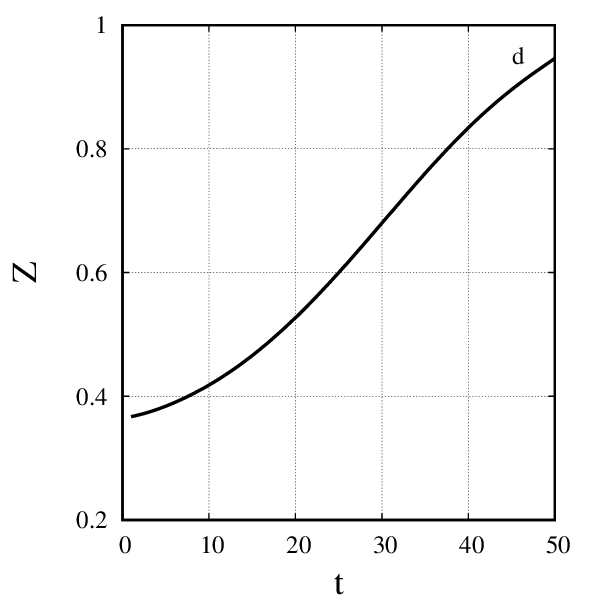} 
	\caption{Dependencies of channel gap on time for different values of thermal noise stress: 
		a -- $\sigma_t=0.05$,  b -- $\sigma_t=0.08$,  c -- $\sigma_t=0.11$, and  d -- $\sigma_t=0.15$. The van der Waals force stress value is $\sigma_V=0.3$.}
	\label{Z_T_ST}
\end{figure}

This picture can be clearly seen from time evolution of the channel gap $Z$. The dependencies of $Z$ on time for different values of thermal noise stress $\sigma_t$ are presented in figure~\ref{Z_T_ST}.

The criterion of a clean channel is $Z=0.95$, which is used in the calculation algorithm. At high values of $\sigma_t>0.1$ the gap value reaches this criterion rather quickly (Fig.~\ref{Z_T_ST}d), which indicates complete cleaning of the channel. At the threshold value $\sigma_t=0.08$, a significant slowdown of the cleaning process is observed. Thus, here the contribution of viscous stresses is rather small, which leads to the actual stopping of the cleaning process. Indeed, for $\sigma_t=0.05$, the gap in general has no clear tendency either to increase or to decrease; there are little noticeable chaotic oscillations caused by thermal fluctuations, and the channel cleaning does not occur.

Another key parameter helping to control the cleaning process is the the van der Waals force stress tensor $\sigma_V$. Let us illustrate the effect of ability of particle attachment on the evolution of the integral parameters for two values of the thermal stress: at low temperature $\sigma_t=0.05$ and at high temperature $\sigma_t=0.1$.

\begin{figure}
	\includegraphics[width=0.49\linewidth]{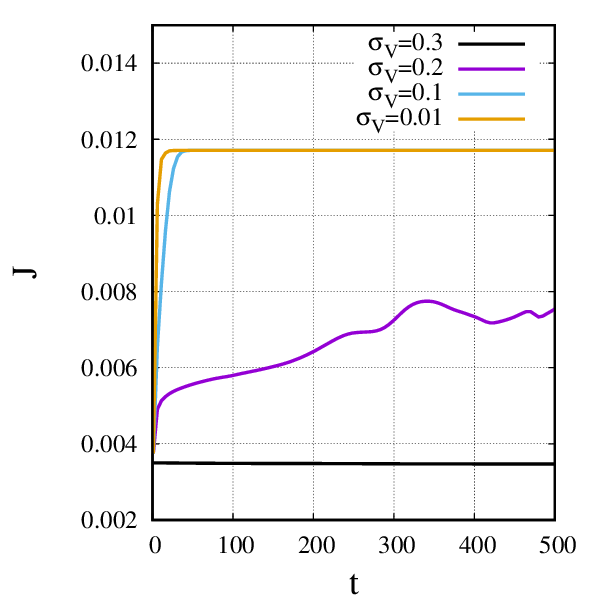} 
	\includegraphics[width=0.49\linewidth]{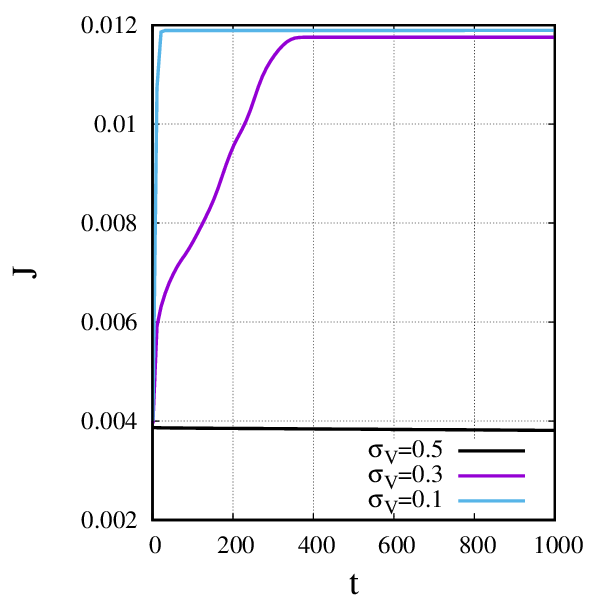} 
	\caption{Dependencies of fluid flux $J$ on time for different values of Van der Vaals force stress $\sigma_V$ which is indicated in legend. The thermal noise stress values are $\sigma_t=0.05$ (top) and $\sigma_t=0.1$ (bottom).}
	\label{J_T_SV}
\end{figure}

Time dependencies of the fluid flux ($J$) through the channel for different values of $\sigma_V$ are plotted in Fig. \ref{J_T_SV}. It is shown, at low temperature at $\sigma_t=0.05$, the critical value of $\sigma_V$ is about $\sigma_V\approx 0.2$, and at higher values of $\sigma_V$ the channel can not be cleared by pure liquid flow. Moreover, at $\sigma_V= 0.2$, the cleaning process is very slow. At high temperature ($\sigma_t=0.1$), the channel cleaning is observed for higher values of the van der Waals force stress, specifically, at $\sigma_V=0.3$. The cleaning process in this case is also not so fast; the time of total cleaning is about $t=350$ dimensionless units. However, at high temperature it is possible to clean the channel, contradictory to the case of low temperature $\sigma_t=0.05$.

\begin{figure}
	\includegraphics[width=0.49\linewidth]{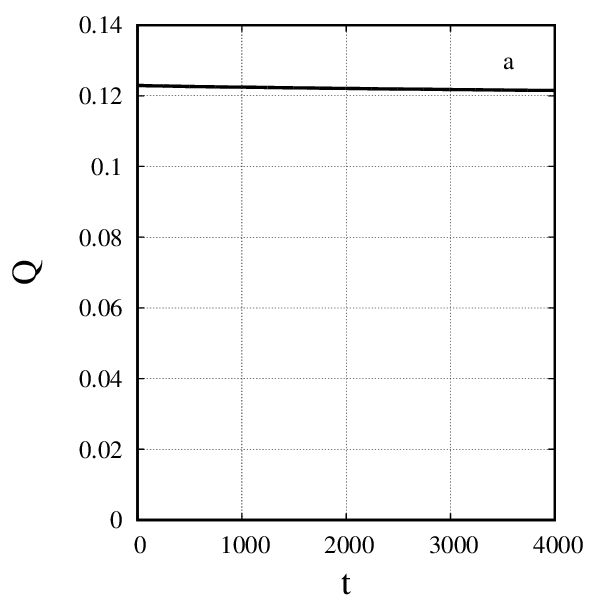} 
	\includegraphics[width=0.49\linewidth]{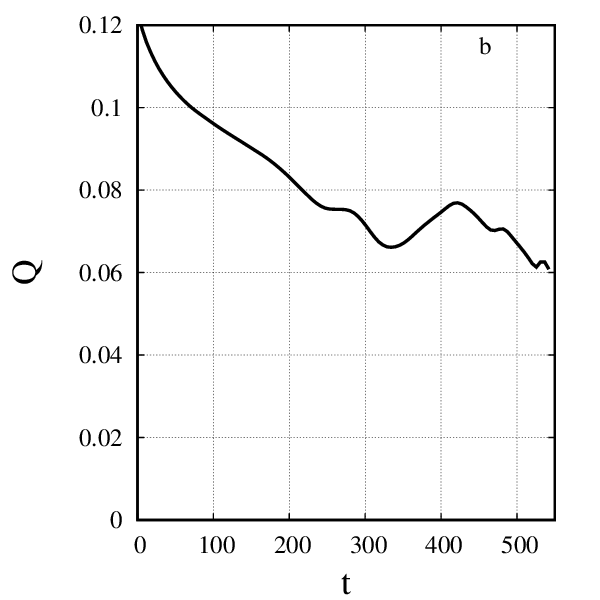} 
	\includegraphics[width=0.49\linewidth]{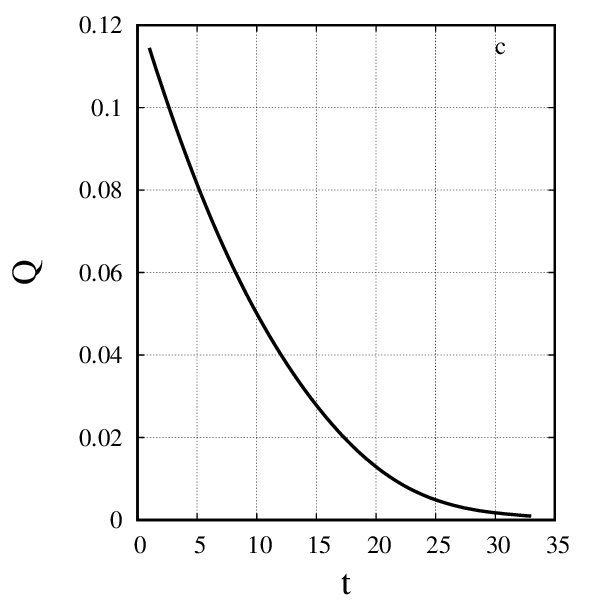} 
	\includegraphics[width=0.49\linewidth]{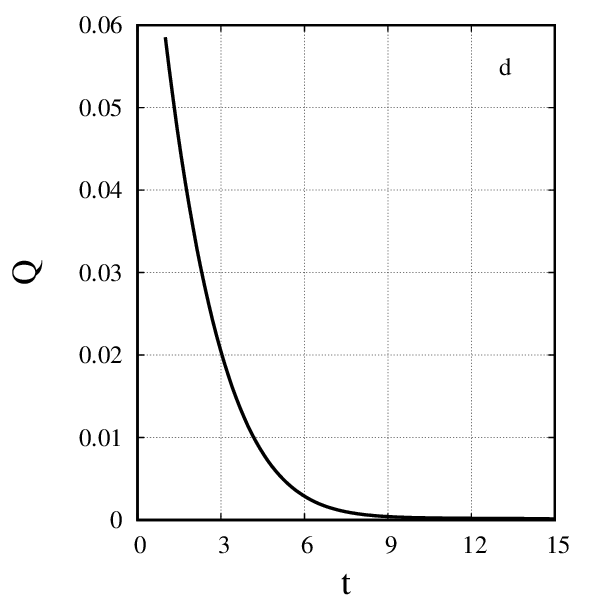} 
	\caption{Dependencies of  occupied volume ($Q$) on time for different values of van der Waals force stress: 
		a -- $\sigma_V=0.3$,  b -- $\sigma_V=0.2$,  c -- $\sigma_V=0.1$, and  d -- $\sigma_V=0.01$. The thermal noise stress value is $\sigma_t=0.05$.}
	\label{Q_T_SV1}
\end{figure}

\begin{figure}
	\includegraphics[width=0.49\linewidth]{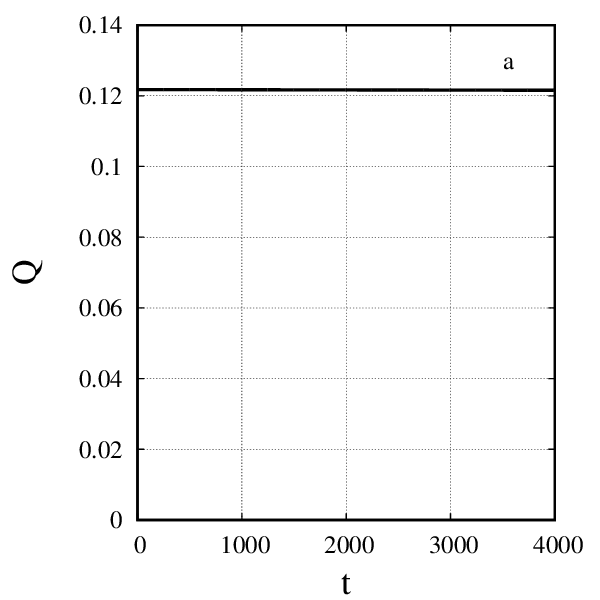} 
	\includegraphics[width=0.49\linewidth]{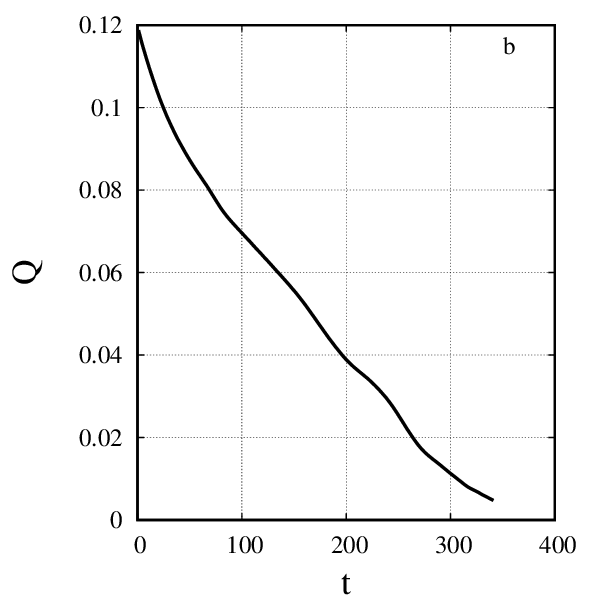} 
	\includegraphics[width=0.49\linewidth]{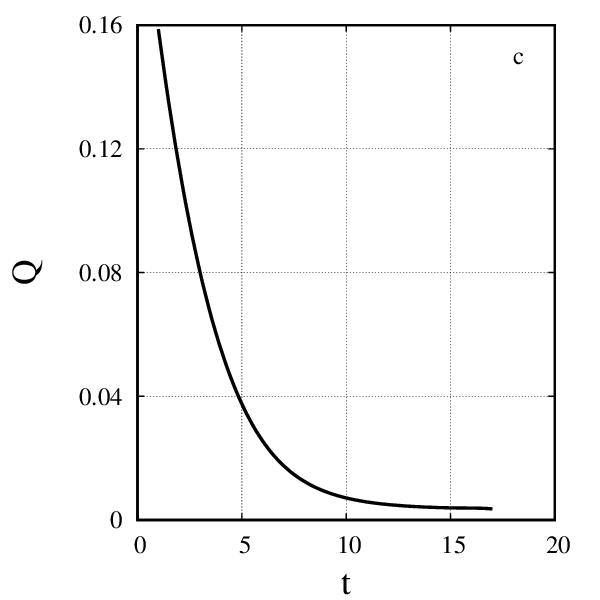} 
	\caption{Dependencies of occupied volume ($Q$) on time for different values of Van der Waals force stress: 
		a -- $\sigma_V=0.5$,  b -- $\sigma_V=0.3$ and  c -- $\sigma_V=0.1$. The thermal noise stress value is $\sigma_t=0.1$.}
	\label{Q_T_SV2}
\end{figure}

Dependencies of the occupied volume concentration $Q$ on time are presented in Fig.~\ref{Q_T_SV1} ($\sigma_t=0.05$) and Fig.~\ref{Q_T_SV2} ($\sigma_t=0.1$). The same picture of the cleaning process, as described above, is demonstrated; with an increase of the thermal noise stress $\sigma_t$, the critical value of $\sigma_V$ is also enlarged.

\begin{figure}
	\includegraphics[width=0.49\linewidth]{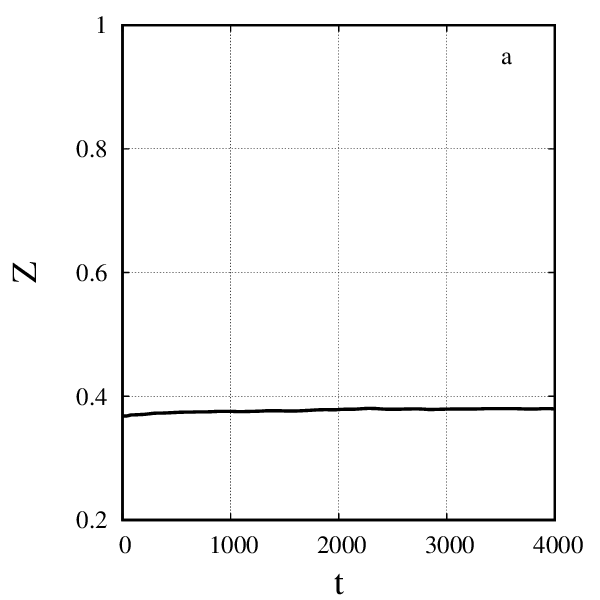} 
	\includegraphics[width=0.49\linewidth]{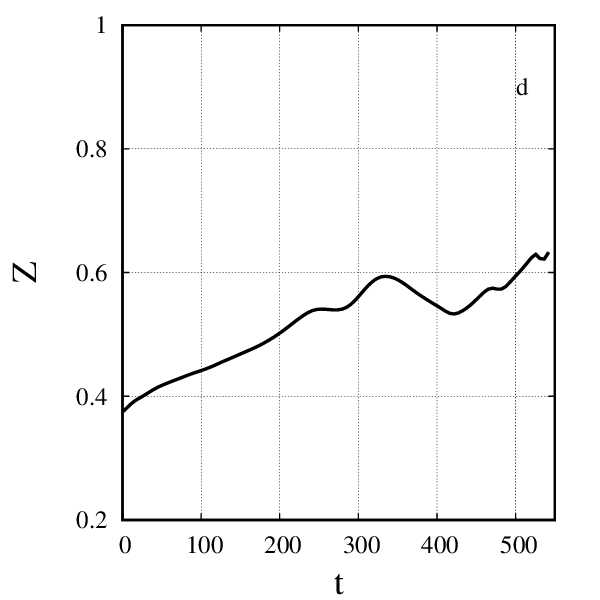} 
	\includegraphics[width=0.49\linewidth]{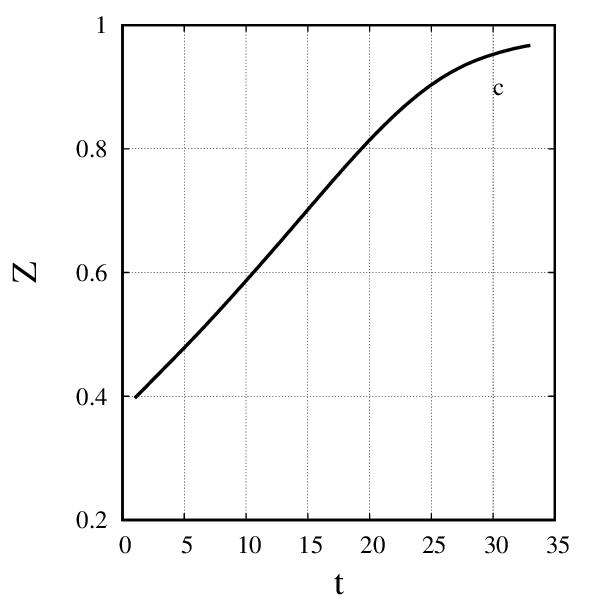} 
	\includegraphics[width=0.49\linewidth]{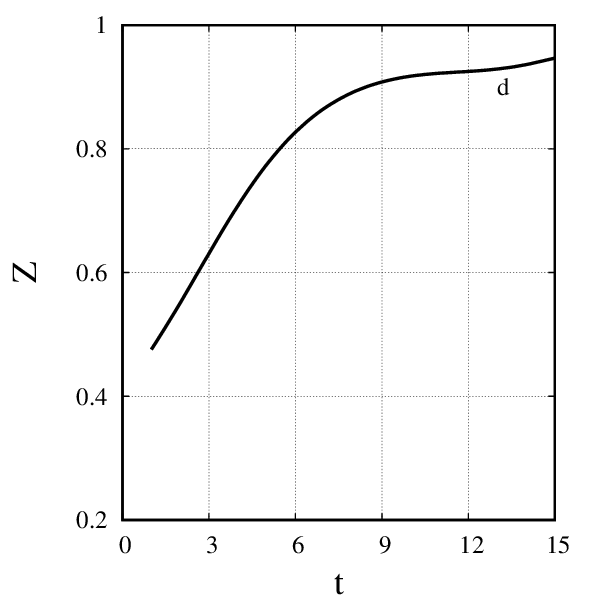} 
	\caption{Dependencies of channel gap ($Z$) on time for different values of the van der Waals force stress: 
		a -- $\sigma_V=0.3$, b -- $\sigma_V=0.2$, c -- $\sigma_V=0.1$ and d -- $\sigma_V=0.01$. The thermal noise stress value is $\sigma_t=0.05$.}
	\label{Z_T_SV1}
\end{figure}

\begin{figure}
	\includegraphics[width=0.49\linewidth]{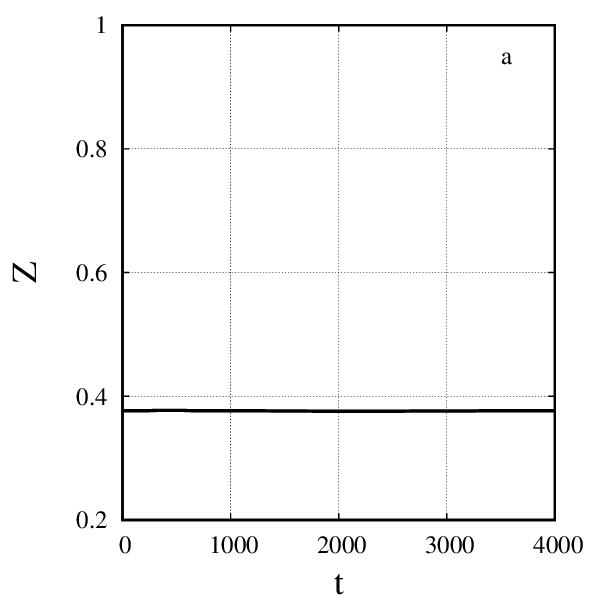} 
	\includegraphics[width=0.49\linewidth]{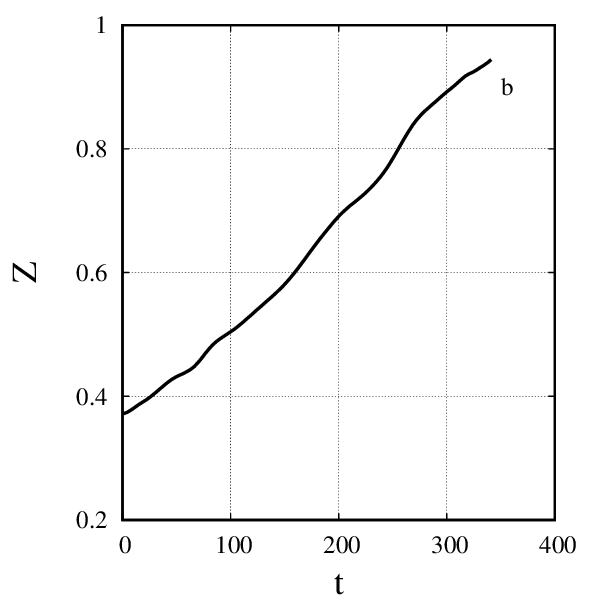} 
	\includegraphics[width=0.49\linewidth]{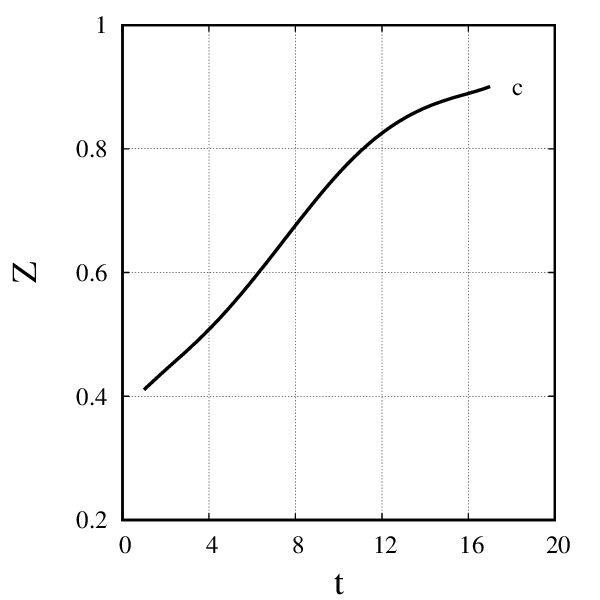} 
	\caption{Dependencies of channel gap ($Z$) on time for different values of the van der Waals force stress: 
		a -- $\sigma_V=0.5$,  b -- $\sigma_V=0.3$ and  c -- $\sigma_V=0.1$. The thermal noise stress value is $\sigma_t=0.1$.}
	\label{Z_T_SV2}
\end{figure}

The dependensies of $Z$ on time show the same processes and they presented in figures Fig.~\ref{Z_T_SV1}, Fig.~\ref{Z_T_SV2}.

Thus, three scenarios of the channel cleaning process can be distinguished. The first one (I) corresponds to the absence of cleaning; the channel remains clogged (Fig.~\ref{Q_T_ST}a, Fig.~\ref{Q_T_SV1}a, Fig.~\ref{Q_T_SV2}a). The intensity of the van der Waals forces is much greater than the sum of viscous and random forces. In this case all integral variables kept near constant values, and the particle detachment cannot be observed. 

In the second scenario (II) the cleaning is slow (Fig.~\ref{Q_T_ST}b, Fig.~\ref{Q_T_SV1}b, Fig.~\ref{Q_T_SV2}b). In this case the intensities of the van der Waals force and the sum of viscous and random forces are comparable. Curves of integral parameters $Q(t)$, $J(t)$, $Z(t)$ have a rather complex form, however trend to the occupied volume $Q$ decrease with time has been detected. In this case the particles are able to detach from the channel wall but quite often they do not go out from the channel and attach to the wall in other place. So, as a result, one can observe complex attachement -- detachement process. 

The last scenario (III) (Fig.~\ref{Q_T_ST}c, Fig.~\ref{Q_T_ST}d Fig.~\ref{Q_T_SV1}c, Fig.~\ref{Q_T_SV1}c and Fig.~\ref{Q_T_SV2}c) corresponds to fast cleaning process. For this type,
the sum of viscous and random forces is greater than the van der Waals force. So one can observe fast decreasing of occupied volume and also fast increasing the channel gap and flux of through flow. Most of the time particles took part in detachment process, while particle attachment to the wall is very rare. It is shown that this regime can be describes mathematically by simple linear desorption expression:

\begin{equation}
	\frac{\partial Q}{\partial t}=-bQ
	\label{desorp}
\end{equation}
where $b$ is linear desorption rate.

Time dependencies of $Z$ are presented in Fig.~\ref{Z_T_SV1}, Fig.~\ref{Z_T_SV2}.

The criteria of calculation stopping is achievement the value $Z=0.95$. If the ratio between stresses ensures the fast cleaning (scenario III), then a linear growth of the channel gap curve is observed. This growth is explained by the fact that particle detachment occurs regularly (each time step). Then the growth of the curve slows down due to the rarer particle detachment.

Thus, we have demonstrated that the scenario of channel cleaning depends on the values of two governed parameters: $\sigma_t$, which is linked to the temperature, and $\sigma_V$, which is the characteristic of the particle and wall material. Further, we will discuss the dependence of cleaning time $T*$ (time when the channel gap reaches the value $Z=0.95$) on these parameters $\sigma_t$ and $\sigma_V$. 

\begin{figure}
	\includegraphics[width=0.49\linewidth]{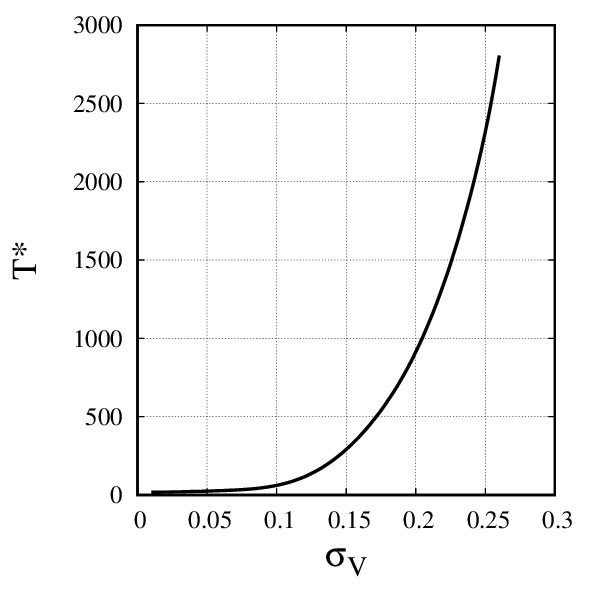} 
	\includegraphics[width=0.49\linewidth]{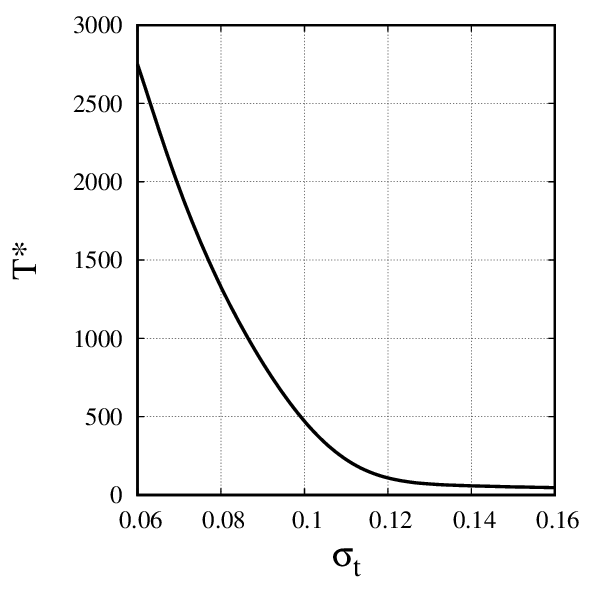} 
	\caption{Dependencies of cleaning time $T*$ on specific stress: on $\sigma_V$ at $\sigma_t=0.05$ (top) and  on  $\sigma_t$ at $\sigma_V=0.3$ (bottom).}
	\label{T_S}
\end{figure}	

As it is shown in Fig.~\ref{T_S}, the cleaning time extremely grows with increasing $\sigma_V$ and decreasing $\sigma_t$. The first parameter, $\sigma_V$, describes the intensity of adhesion between the channel wall and particles of admixture. The van der Waals force stress $\sigma_V$ is defined by formula (\ref{str_V}) and depends on the Hamacker constant $A$, particle size $a$, and pressure drop between the channel inlet and outlet $\Delta P$. The first parameter, $A$, characterizes the dielectric properties of the admixture and channel wall, which is the essence of the parameter $\sigma_V$. Two other parameters, $a$ and $\Delta P$, characterize the problem as a whole and were introduced as dimensionless units. Its variations lead to the variations of all dimensionless parameters of the problem. Parameter $\sigma_t$ describes the random force that is generated by the thermal noise. This parameter is mainly defined by temperature (see (\ref{str_T})), while other variables in (\ref{str_T}) are also introduced for dimensionalization. The curves in Fig.~\ref{T_S} demonstrate the balance between attachment and detachment processes that characterizes the cleaning time. 

\begin{figure}
	\includegraphics[width=0.49\linewidth]{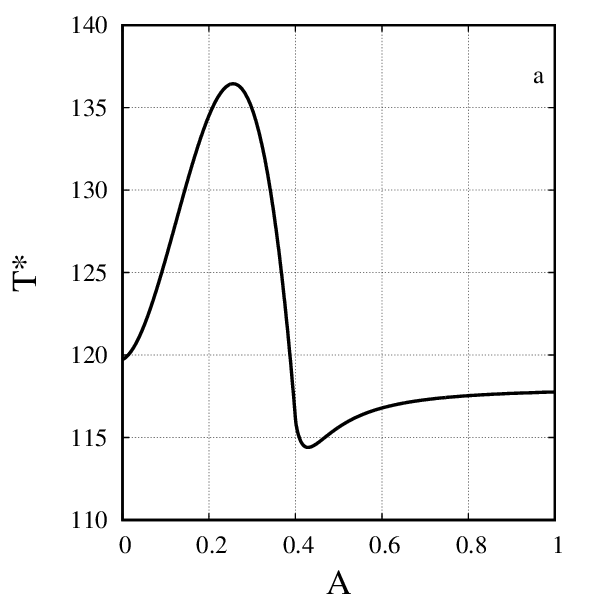} 
	\includegraphics[width=0.49\linewidth]{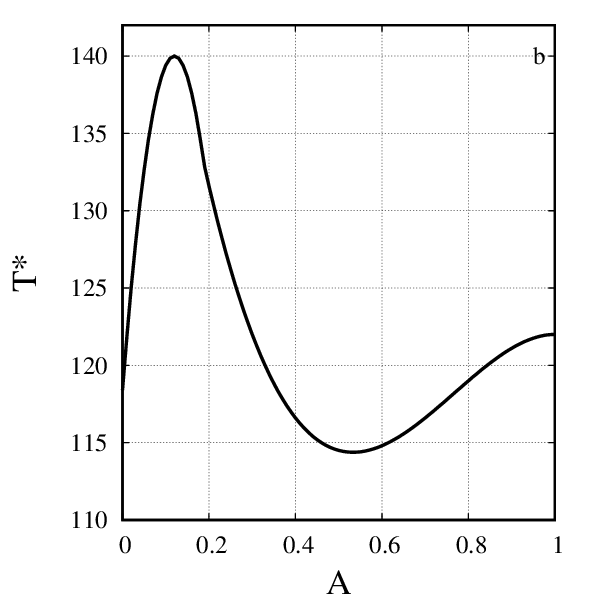} 
	\includegraphics[width=0.49\linewidth]{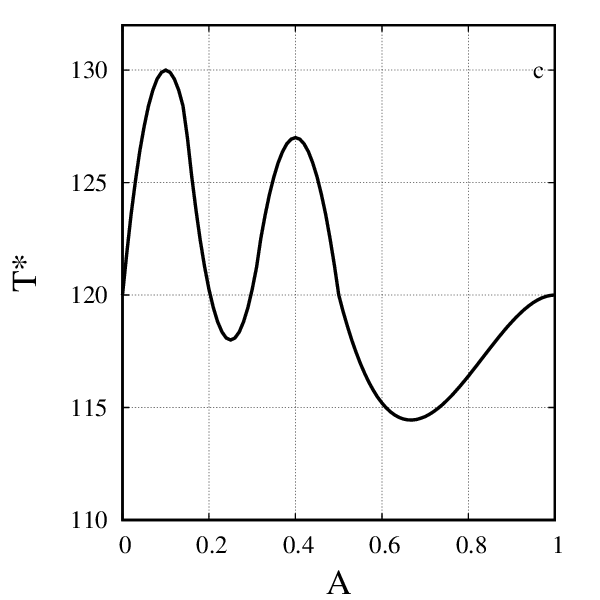} 
	\caption{Dependencies of cleaning time ($T*$) on amplitude of modulation $A$ for different values of modulation frequency: 
		a -- $\omega=0.01$,  b -- $\omega=0.04$ and  c -- $\omega=0.07$. The stress values are $\sigma_V=0.15$ and $\sigma_t=0.1$.}
	\label{T_A}
\end{figure}

One of the simplest ways of controlling the cleaning process is to use a time-modulated flow of pure liquid instead of a stationary one. The equations (\ref{Gover_mod}) have been solved for this purpose. The main interest is in studying the effect of modulation on the cleaning time $T*$. The change in the relative modulation amplitude is carried out in the interval $A\in[0,1]$, which allows maintaining the direction of the flow from the inlet to the outlet of the channel. The modulation frequency is $\omega\ll 1$, which follows from the fact that in microfluidic devices the modulation period should be less than the mean free time of particles. Fig.~\ref{T_A} shows the dependences of $T*$ on the modulation amplitude of the flow $A$.

From Fig.~\ref{T_A} it is evident that the clearing time without modulation is about 120 dimensionless units. For different amplitude values the clearing time can be both shorter and longer than the clearing time without modulation. The number of such intervals increases with increasing frequency. In the case of $\omega=0.01$ there is only one interval with a longer time at $A\in [0,0.38]$ and one with a shorter time $A\in [0.38,1]$. At a higher frequency ($\omega=0.07$) one can observe two amplitude intervals with a longer clearing time ($A\in[0,0.19]$ and $A\in[0.28,0.52]$ and two with accelerated clearing ($A\in[0.19,0.28]$ and $A\in[0.52,1]$). One can also see local maxima and minima on the $T*(A)$ dependencies.

\begin{figure}
	\includegraphics[width=0.45\linewidth]{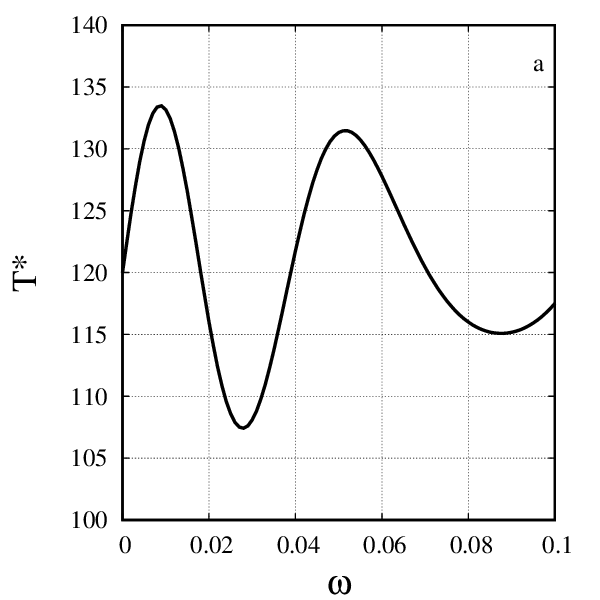} 
	\includegraphics[width=0.45\linewidth]{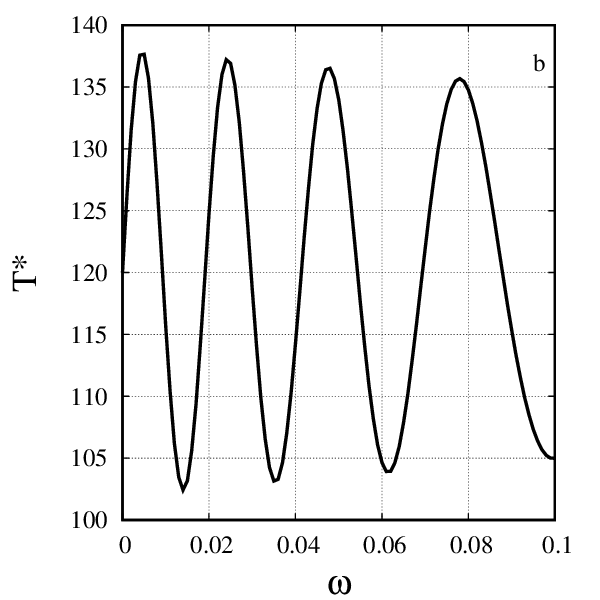} 
	\includegraphics[width=0.45\linewidth]{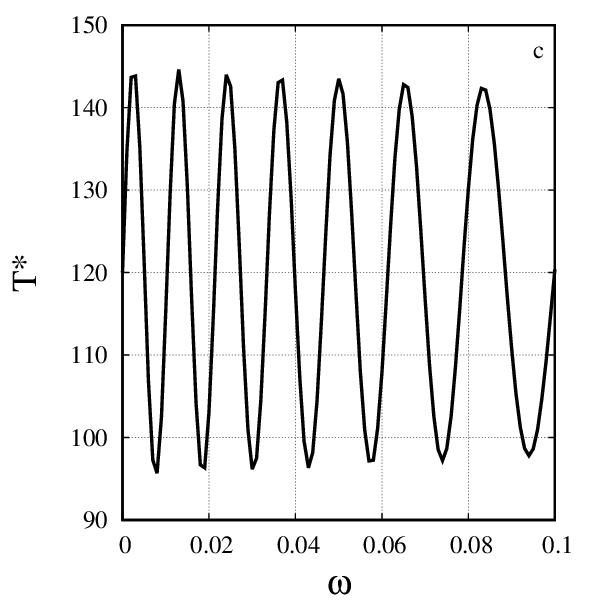} 
	\caption{Dependencies of cleaning time ($T*$) on frequency of modulation $\omega$ for different values of modulation amplitude: 
		a -- $A=0.3$, b -- $A=0.5$ and c -- $A=0.8$. The  stress values are $\sigma_V=0.15$ and $\sigma_t=0.1$.}
	\label{T_W}
\end{figure}

In our opinion, all these features indicate the presence of resonance phenomena in this system. Indeed, let us turn to the second cleaning scenario with numerous attachments and detachments. In this case, very slow cleaning is observed, $T*>100$, and it is obvious that there is some average time between the attachment and detachment of particles. Thus, these two processes are periodic. The period of such processes is significantly greater than the dimensionless unit of time; if the period of the modulated flow coincides with the period of particle detachment, the cleaning process accelerates. Otherwise, the cleaning process slows down. The number of such time intervals is expected to increase with increasing frequency.  Fig.~\ref{T_W} presents the dependence of the cleaning time $T$ on the modulation frequency  $\omega$ for different values of the amplitude $A$.

The shape of the curves in Fig.~\ref{T_W} is similar to oscillations with a change in amplitude. This shape is also the result of a resonant process. The amplitude of the change in the cleaning time increases with increasing pressure, since the acceleration and deceleration of the flow are proportional to the pressure. All other characteristics are similar to the previous results in Fig.~\ref{T_A}. The presence of resonance allows us to control the cleaning process by changing the amplitude and frequency of the pressure modulation.

\section{Conclusion}
\label{Conc}

The process of cleaning a microchannel contaminated by impurity particles deposited on its walls is studied. The cleaning is caused by the pure fluid throughflow at a given pressure drop between the inlet and outlet of the channel. The problem is solved in a two-dimensional statement under the assumptions of laminar flow, the absence of inertial effects for contaminant particles, and van der Waals interactions between the particles and the wall. A mathematical model and computational algorithm have been developed. The model takes into account the viscous force from the flow and the random force, caused by the thermal fluctuations. The problem is solved numerically in the framework of the random walk model with statistical processing of obtaining data. An example of a single realization of the evolution of the channel shape during cleaning is presented. Depending on the problem parameters, a typical picture of the microchannel cleaning is the following: the initially clogged microchannel quite fast becomes clean everywhere except for the largest particle deposits. However, the complete cleaning can be achieved in some parameter range; otherwise, the channel remains clogged.

Statistical processing of at least 50 realizations leads to an analysis of the evolution of integral variables. The dependences of fluid flow through the channel, the volume occupied by stuck particles, and the channel gap on time have been analyzed. It is shown that there are two governed parameters, which control the cleaning process, specifically: the random thermal stress and the van der Waals force stress. Their ratio determines the observed cleaning scenario. Three possible scenarios can be observed: no cleaning, slow cleaning, and fast cleaning.

The possibility of controlling the cleaning process by modulating in time the fluid flow through the channel was investigated. Resonance phenomena were observed. The dependence of cleaning time on amplitude and frequency was obtained and analyzed. It was shown that changing the modulation parameters leads to a significant change in cleaning time, enabling control of the cleaning process.

\backmatter

\bmhead{Acknowledgments}
The work was supported by the Russian Science Foundation
(grant No. 23 - 12 - 00180 ).

\bmhead{Author contributions}
Boris Maryshev (first author, corresponding author) Conceptualization,  Resources, Supervision, Methodology, Investigation, Writing Original Draft, Funding Acquisition. Lyudmila Klimenko (second author) Investigation, Formal Analysis, Text Correction, Visualization, Numerical calcullations.  All
authors reviewed the manuscript. 

\bmhead{Funding} The work was supported by the Russian Science Foundation (grant No. 23 - 12 - 00180).

\bmhead{Data Availability} All data included in this study are available into the paper or upon request by contact with the corresponding author

\section*{Declarations}
\bmhead{Submition} The materials of this preprint where submited to Microgravity Science and Technology journal.
\bmhead{Ethical Approval} The authors confirm that this work is original and has not been published elsewhere nor is it currently under consideration for publication elsewhere.
\bmhead{Consent to Participate} The authors declare that they all participated in this study.
\bmhead{Consent for Publication} The authors declare that they all agree to the publication of this paper.
\bmhead{Competing Interests} The authors declare that they have no competing financial interests

\bibliography{bibliography}

\end{document}